\documentclass[11pt,a4paper]{article}
\usepackage{jheppub}
\usepackage{amsfonts}

\title{Correlation Functions in N=3 Superconformal Theory}
\author[a,1]{Dmitriy Drichel,%
\note{Corresponding author}}
\author[a]{Michael Flohr}

\affiliation[a]{Institute for Theoretical Physics,\\
 Appelstra\ss e 2, 30167 Hannover, Germany}
\emailAdd{dmitriy.drichel@itp.uni-hannover.de}
\emailAdd{michael.flohr@itp.uni-hannover.de}
\abstract{Using a superspace representation of the N=3 Neveau-Schwarz super Virasoro algebra, we find solutions of N=3 super Ward identities. Global transformations generated by the non-abelian supercurrent require not only superfields, but also functions of Grassmann variables (in particular correlation functions and their linear combinations)  to be su(2) representations. As a consequence, the only admissible fields in the theory are isospin singlets and doublets. We show how to compute the generic form of any N=3 $n$-point function and demonstrate a construction of all su(2) representations on the space of N=3 superfunctions.}
\arxivnumber{1006.3346}
\begin{document}

\maketitle
\flushbottom

\section{Introduction}
\noindent Superconformal extensions of the Virasoro algebra  have been studied in the context of high-energy physics \cite{Boucher:1986bh}\cite{Greene:1996cy}, two-dimensional critical systems \cite{Qiu:1986if} and as a subject on its own  \cite{SchwimmerSeiberg}\cite{Dorrzapf:1997rx}\cite{Nagi:2003pb}\cite{Nagi:2004wb}.

The N=1 and N=2 theories remain the most studied and well-understood super Virasoro theories. Supersymmetric extensions for different N turn out to have surprisingly distinctive properties. Some of the features of the N=2 theories include existence of subsingular vectors \cite{GatoRivera:1996tc} and unitarity of all rational theories \cite{Eholzer:1996zi}. Central extensions of the superconformal algebra do not exist for N$>$4 \cite{Schoutens:1988ig}.

In this paper, the $n$-point functions in the Neveau-Schwarz sector of N=3 super Virasoro theories, as far as they are fixed by global transformations, are studied. 

We find that the superfield representation space of the non-abelian supercurrent is restricted to the lowest two representations. This allows us to calculate arbitrary $n$-point correlation functions. In fact, our results indicate that the representation content of N$>$3 theories with respect to the supercurrent is finite-dimensional as well, which would make N=2 the most ``interesting'' super Virasoro theory with the largest representation content.  

Calculation of correlation functions in supersymmetric extensions with N$>$2 involves non-abelian symmetry generators, returning systems of superdifferential equations containing correlation functions of fields in different states with respect to the Cartan basis of the supersymmetry current. The super Ward identities for N=3 lead to equations containing correlation functions of fields in a representation $t$ with raised or lowered eigenvalues $\vert q \vert \leq t$ of the su(2) generator $T_{0}^{H}$. The situation is different in N=2, where $q$ is an eigenvalue of a U(1) supercurrent, the correlation functions are restricted by charge conservation and the Ward identities return differential equations in only one correlation function. We will refer to $q$ as isospin.

In section 2 and 3 we fix the notation by introducing N=3 superconformal transformations of superfunctions and superprimary fields, respectively. In section 4, we determine the two-point correlation functions by solving super Ward identities. The general $n$-point function is produced by contraction in section 5. We discuss implications of our findings in the final part of this manuscript.

\section{Global Transformations in N=3 Superspace}

\noindent We consider the two-dimensional N=3 superspace with a basis $\mathsf{Z}$ given by a complex coordinate $z$ and three Grassmannian variables $\theta^{1}$, $\theta^{2}$ and $\theta^{3}$.
The superconformal condition $\omega'=\kappa\left(z,\theta_{i}\right)\omega$ for the one-form 
$\omega=dz-\sum_{i}d\theta_{i}\theta_{i}$ admits a set of classical generators of superconformal transformations (using integer $m$, half-integer $r$ and factors $\frac{1}{2}$ in hindsight)
\begin{equation}\label{cl}
\begin{array}{rcl}
 l_{m}&=&-z^{m}\left(z\partial_{z}+\frac{1}{2}\left(m+1\right)\theta^{i}\partial_{\theta^{i}}\right),\\
 g_{r}^{i}&=&\phantom{-}\frac{1}{2}z^{r-\frac{1}{2}}\left(z\theta^{i}\partial_{z}-z\partial_{\theta^{i}}+\left(r+\frac{1}{2}\right)\theta^{i}\theta^{j}\partial_{\theta^{j}}\right),\\
 t^{i}_{m}&=&\phantom{-}\frac{1}{2}z^{m-1}\left(z\epsilon_{ijk}\theta^{j}\partial_{\theta^{k}}-m\theta^{1}\theta^{2}\theta^{3}\partial_{\theta^{i}}\right),\\
 \psi_{r}&=&-\frac{1}{2}z^{r-\frac{1}{2}}\left(\theta^{1}\theta^{2}\theta^{3}\partial_{z}+\frac{1}{2}\epsilon_{ijk}\theta^{i}\theta^{j}\partial_{\theta^{k}}\right).
\end{array}
\end{equation} 
The (anti-)commutation relations of the global subset of these generators are a representation of the classical N=3 superconformal algebra.
The subset of global transformations is the set of all generators that has no singularities on the bosonic coordinate of a graded Riemann sphere \cite{Nagi:2003pb}. This subset consists of twelve generators of the group OSp(2$\vert$3),
\begin{equation*}
\lbrace l_{\pm 1,0}, g_{\pm \frac{1}{2}}^{i}, t_{0}^{i} \rbrace. 
\end{equation*}
The operators $ l_{\pm 1,0}$ are simply an extension of the conformal N=0 group to superspace, $g_{- \frac{1}{2}}^{i}$ generate supertranslations and $g_{\frac{1}{2}}^{i}$ are generators of special superconformal transformations. The closed set of operators $t_{0}^{i}$ is a manifestation of the so(3) symmetry of fermionic coordinates which leaves $z$ unaffected.

In contrast to continous rotations in Euclidian spaces, transformations generated by so(N) operators are discrete on fermionic coordinates. The supercurrent is diagonalized by mapping its modes $t_{m}^{i}$ to an su(2) basis,
\begin{equation*}
\begin{array}{rclcrclcrcl}
 \theta^{+}&=&2\left(i\theta^{1}-\theta^{2}\right), &\phantom{mm} &
 \theta^{-}&=&2\left(i\theta^{1}+\theta^{2}\right), &\phantom{mm} &
 \theta^{H}&=&i\theta^{3},\\
 t^{+}_{m}&=&2\left(it^{1}_{m}-t^{2}_{m}\right), & &
 t^{-}_{m}&=&2\left(it^{1}_{m}+t^{2}_{m}\right), & &
 t_{m}^{H}&=&-2it^{3}_{m}, \\
 g_{r}^{+}&=&4\left(g_{r}^{2}-ig_{r}^{1}\right), & &
 g_{r}^{-}&=&4\left(g_{r}^{2}+ig_{r}^{1}\right), & &
 g_{r}^{H}&=&8ig_{r}^{3}.
\end{array}
\end{equation*}

Because of the discrete nature of their action on nilpotent coordinates, the su(2) generators $t_{0}^{\pm}$ satisfy $ \left(t_{0}^{\pm}\right)^{3}=0 $. As a consequence of this, the highest superfunction representation in N=3 Grassmannian variables  is the $t=1$ representation. In fact, the only admissible superfunction representations are $t=0$ and $t=1$, which can be seen by considering coefficients in the expansion of a superfunction with respect to nilpotent varialbles,
\[
 \lbrace 1, \theta^{H}, \theta^{+}, \theta^{-},\theta^{H}\theta^{+}, \theta^{+}\theta^{-}, \theta^{H}\theta^{-}, \theta^{H}\theta^{+}\theta^{-}\rbrace,
\]
and calculating all elements through the generators of su(2),
\[
\begin{array}{ccc}
 t_{0}^{H}=\theta^{+}\partial_{\theta^{+}}-\theta^{-}\partial_{\theta^{-}}
 & \mathrm{and} &
t_{0}^{\pm}=\pm\frac{1}{2}\theta^{\pm}\partial_{\theta^{H}}\mp 4\theta^{H}\partial_{\theta^{\mp}}.
\end{array}
\]

The terms in the expansion provide a basis for constructing $t=0$ and $t=1$ representations in superspace,
\begin{equation}\label{su2basis}
 \begin{array}{rl}
 t=0,\,\ q=\phantom{-}0: & \lbrace 1, \theta^{+}\theta^{-}\theta^{H} \rbrace, \\
 t=1,\,\ q=\phantom{-}1: & \lbrace \theta^{+}, \theta^{+}\theta^{H} \rbrace, \\
 t=1,\,\ q=\phantom{-}0: & \lbrace \theta^{H}, \theta^{+}\theta^{-} \rbrace, \\
 t=1,\,\ q=-1: & \lbrace \theta^{-}, \theta^{-}\theta^{H} \rbrace. \\
 \end{array}
\end{equation}

We can not draw any conclusions about superfield representations from this result alone, since superfields are not guaranteed to have a superspace representation. A primary superfield written using the notation $\Phi_{h,t,q}(z, \theta^{+}, \theta^{-}, \theta^{H})$ should not be mistaken for a genuine superfunction. It can not, in general, be expanded in Grassmannian variables. However, correlation functions of superfields are ordinary functions dependent on superspace variables, and this will allow us to draw conclusions about the representation space of conformal superfields. 

\section{Superfields in N=3 Theory}
\noindent An immediate consequence of the discussion in the previous section is that OSp(2$\vert$3) is the largest subgroup of transformations that annihilates the conformal N=3 vacuum.
 To obtain a quantum algebra of infinitesimal transformations of superconformal quantum field theory, the algebra of classical generators is centrally extended. Consistency with the super Jacobi identity leads to the complete quantum N=3 Neveau-Schwarz algebra
\begin{equation*}
\begin{array}{rclrcl}
 [L_{m}, L_{n}]&=&\left(m-n\right)L_{m+n}+\frac{k}{4}m\left(m^{2}-1\right)\delta_{m+n,0}, \;\;\;
 & [L_{m}, T_{n}^{\pm,H}]&=&-n T_{m+n}^{\pm,H} ,
\\

 \lbrace G_{r}^{H}, G_{s}^{H} \rbrace &=& -32L_{r+s}-16k \left( r^{2}-\frac{1}{4}\right)\delta_{r+s,0}, 
& {}[L_{m}, G_{r}^{\pm,H}]&=&
\left( \frac{m}{2}-r\right) G_{r+m}^{\pm, H},
\\
\lbrace G_{r}^{+}, G_{s}^{-} \rbrace &=& 16L_{r+s}+8k\left(r^{2}-\frac{1}{4}\right)\delta_{r+s,0}
 &  {}[L_{m}, \psi_{s}]&=&-\left(\frac{m}{2}+s\right)\psi_{m+s},
\\
& &+8\left( r-s \right)T^{H}_{r+s}, 
 & [T^{\mp}_{m}, G_{r}^{\pm}]&=&-G^{H}_{r+m}\pm 8m \psi_{r+m}, \\
 \lbrace G_{r}^{\pm}, G_{s}^{H} \rbrace &=& 8\left(r-s \right)T^{\pm}_{r+s},	
&
[T^{\pm}_{m}, G^{H}_{r}]&=&-2G_{m+r}^{\pm}, \\
 {}[T^{+}_{m}, T^{-}_{n}]&=&2T^{H}_{m+n}+2km\delta_{m+n,0},
&
  {}[T_{m}^{H}, G_{r}^{\pm}]&=&\pm G_{r+m}^{\pm},\\
{}[T_{n}^{H}, T_{m}^{H}]&=&km\delta_{m+n,0},
&
 [T^{H}_{m}, G_{r}^{H}]&=&-2T^{H}_{r+s},\\
{}[T_{m}^{H}, T_{n}^{\pm}]&=&\pm T^{\pm}_{m+n},  
&
 \lbrace \psi_{s}, G_{r}^{\pm} \rbrace&=&\mp T^{\pm}_{r+s}, \\
{}[\psi_{r}, \psi_{s}]&=&-\frac{k}{4}\delta_{r+s,0},
& \lbrace \psi_{s}, G^{H}_{r} \rbrace &=&-2T_{r+s}^{H}, 


\end{array}
\end{equation*}
with all other (anti-)commutator relations vanishing.

The Cartan subalgebra is two-dimensional. A highest-weight state is simultaneously diagonalizable in its conformal weight and isospin,
\begin{equation}\label{hw}
\begin{array}{cc}
L_{0} \vert h,q\rangle =h \vert h, q\rangle, \;\;\; & T^{H}_{0}\vert h, q \rangle = q\vert h,q \rangle.
\end{array}
\end{equation}
We further expect the highest-weight states to satisfy
\begin{equation*}
L_{n}\vert h, q \rangle = T^{\pm,H}_{n} \vert h, q \rangle = G_{\pm r}^{\pm, H} \vert h, q \rangle =\psi_{\pm r} \vert h, q \rangle=0, \;\;\; n,\; r > 0 . 
\end{equation*}

The action of generators of transformations can be given by introducing algebraic su(2) operators $J^{\pm}$, $J^{H}$ which account for the transformation of index $q$ carried by a primary field
\begin{equation}\label{j}
\begin{array}{cc}
 [J^{H}, J^{\pm}]=\pm J^{\pm}, \;\;&\;\;
 [J^{+}, J^{-}]= 2 J^{H}.
\end{array}
\end{equation}
These operators are defined to have following relations with the underlying so(3) generators $J^{i}$
\begin{equation}\label{jj}
\begin{array}{ccc}
J^{+} = 2(iJ^{1} - J^{2}), \;\;&\;\;
J^{-} = 2(i J^{1} + J^{2}), \;\;&\;\;
J^{H} = -2iJ^{3}.
\end{array}
\end{equation}
We choose the convenient normalization
\begin{equation}\label{norm} 
\begin{array}{rcl}
J^{H}\Phi_{h,j,q}(\mathsf{Z})&=&q\Phi_{h,t,q}(\mathsf{Z}),\\
J^{\pm}\Phi_{h,j,q}(\mathsf{Z})&=&\Phi_{h,t,q\pm 1}(\mathsf{Z}).
\end{array}
\end{equation}

The highest-weight conditions on representation spaces of  $L_{0}$ and $J_{0}$ have been treated on the same footing in the N=2 theory, where $q$ is a U(1)-isospin \cite{Kiritsis:1987np}. In contrast to N=2, the isospin representation in N=3 theory is finite-dimensional, since $q$ can take only integer or half-integer values due to (\ref{j}). For every highest-weight state  $\vert h, t, q \rangle$, there is a lowest-weight state $\vert h, t, -q \rangle$. We therefore reserve the term ``primary'' for fields that satisfy (\ref{hw}) without imposing conditions $T^{\pm}_{0}\vert h,\pm q\rangle =0$. Then the infinitesimal transformations of primary fields read
\begin{eqnarray}
\label{L-1}
 [L_{-1}, \Phi\left(\mathsf{Z}\right)]\ =\ \mathcal{L}_{-1}\Phi\left(\mathsf{Z}\right)&=&\partial_{z}\Phi\left(\mathsf{Z}\right), \\
\label{L0}
 [L_{0}, \Phi\left(\mathsf{Z}\right)]\ =\ \ \;\mathcal{L}_{0}\Phi\left(\mathsf{Z}\right)&=&\Big(h+z\partial_{z}+\frac{1}{2}\left(\theta^{+}\partial_{\theta^{+}}+\theta^{-}\partial_{\theta^{-}}+\theta^{H}\partial_{\theta^{H}}\right)\Big)\Phi\left(\mathsf{Z}\right), \\
\label{L1}
 [L_{1},\Phi\left(\mathsf{Z}\right)]\ =\ \ \;\mathcal{L}_{1}\Phi\left(\mathsf{Z}\right)&=&\Big(2hz+z\left(z\partial_{z}+\theta^{+}\partial_{\theta^{+}}+\theta^{-}\partial_{\theta^{-}}+\theta^{H}\partial_{\theta^{H}} \right) \nonumber\\   
 & & \phantom{.}+
 \frac{1}{8}\theta^{+}\theta^{-}J^{H}+\frac{1}{4}\theta^{+}\theta^{H}J^{-}-\frac{1}{4}\theta^{-}\theta^{H}J^{+}\Big)\Phi\left(\mathsf{Z}\right), \\
\label{G-12H}
 [G_{-\frac{1}{2}}^{H}, \Phi\left(\mathsf{Z}\right)]\ =\ \mathcal{G}_{-\frac{1}{2}}^{H}\Phi\left(\mathsf{Z}\right)&=&-4\left(\theta^{H}\partial_{z}+\partial_{\theta^{H}}\right)\Phi\left(\mathsf{Z}\right),\\
\label{G12H}
 [G_{\frac{1}{2}}^{H}, \Phi\left(\mathsf{Z}\right)]\ =\ \ \,\mathcal{G}_{\frac{1}{2}}^{H}\Phi\left(\mathsf{Z}\right)&=&\Big(-8h\theta^{H}-4\theta^{H}z\partial_{z}-4z\partial_{\theta^{H}}-4\theta^{H}\theta^{-}\partial_{\theta^{-}} \nonumber\\
 & & \phantom{.}-
 4\theta^{H}\theta^{+}\partial_{\theta^{+}}+\theta^{-}J^{+}-\theta^{+}J^{-}\Big)\Phi\left(\mathsf{Z}\right), \\
\label{G-12PM}
 [G_{-\frac{1}{2}}^{\pm}, \Phi\left(\mathsf{Z}\right)]\ =\ \mathcal{G}_{-\frac{1}{2}}^{\pm}\Phi\left(\mathsf{Z}\right)&=&\pm\left(\theta^{\pm}\partial_{z}+8\partial_{\theta^{\mp}}\right)\Phi\left(\mathsf{Z}\right), \\
\label{G12PM}
 [G_{\frac{1}{2}}^{\pm}, \Phi\left(\mathsf{Z}\right)]\ =\ \ \,\mathcal{G}_{\frac{1}{2}}^{\pm}\Phi\left(\mathsf{Z}\right)&=&\Big(\pm 2h\theta^{\pm}\pm\theta^{\pm}z\partial_{z}\pm 8z\partial_{\theta^{\mp}}+\theta^{+}\theta^{-}\partial_{\theta^{\mp}}  \nonumber\\
 & & \phantom{.}+
 2\theta^{H}J^{\pm}+\theta^{\pm}J^{H}\Big)\Phi\left(\mathsf{Z}\right),\\
\label{THPHI}
 [T_{0}^{H}, \Phi\left(\mathsf{Z}\right)]\ =\ \ \mathcal{T}_{0}^{H}\Phi\left(\mathsf{Z}\right)&=&\left(\theta^{-}\partial_{\theta^{-}}-\theta^{+}\partial_{\theta^{+}}+J^{H}\right)\Phi\left(\mathsf{Z}\right),\\
\label{T0PM}
 [T_{0}^{\pm}, \Phi\left(\mathsf{Z}\right)]\ =\ \ \mathcal{T}_{0}^{\pm}\Phi\left(\mathsf{Z}\right)&=&\Big(\mp\frac{1}{2}\theta^{\pm}\partial_{\theta^{H}}\pm 4\theta^{H}\partial_{\theta^{\mp}}+J^{\pm}\Big)\Phi\left(\mathsf{Z}\right).
\end{eqnarray}
These transformations are a representation of the superconformal algebra in N=3 superspace and can be used in the form of Ward identities to impose conditions on $n$-point superfunctions. Due to nilpotency of $\theta^{\pm,H}$ and the su(2) nature of isospin, the space of solutions turns out to be surprisingly small. 

\section{Ward Identities and the Two-Point Functions}
\noindent We will use upper indices in parenthesis on superdifferential operators in (\ref{cl}) and (\ref{L-1})-(\ref{T0PM}) for indicating superspace positions $\lbrace z_{i} , \theta^{+}_{i}, \theta^{-}_{i}, \theta^{H}_{i} \rbrace$ on which the operator acts. Then the super Ward identities can be written as sums of one of the differential operators at different superpoints acting on an $n$-point function $F_{n}$,
\begin{eqnarray}
\displaystyle \sum_{i}^{n}\mathcal{L}_{\pm 1,0}^{\left(i\right)}F_{n}\left(\mathsf{Z}_{1}, ... , \mathsf{Z}_{n}\right) &=&  \mathcal{L}_{\pm 1,0}^{\left(1,...,n\right)}F_{n}\left(\mathsf{Z}_{1}, ... , \mathsf{Z}_{n}\right) =0,
\nonumber\\
\displaystyle \sum_{i}^{n}\mathcal{G}_{\pm \frac{1}{2}}^{\pm, H \left(i\right)} F_{n}\left(\mathsf{Z}_{1}, ... , \mathsf{Z}_{n}\right)&=& \mathcal{G}_{\pm \frac{1}{2}}^{\pm, H \left(1,...,n\right)}F_{n}\left(\mathsf{Z}_{1}, ... , \mathsf{Z}_{n} \right)=0,
\nonumber\\
\displaystyle \sum_{i}^{n}\mathcal{T}_{0}^{\pm,H \left(i\right)}F_{n}\left(\mathsf{Z}_{1}, ... , \mathsf{Z}_{n}\right) &=& \mathcal{T}_{0}^{\pm,H \left(1,..,n\right)}F_{n}\left(\mathsf{Z}_{1}, ... , \mathsf{Z}_{n}\right)=0.
\end{eqnarray}
The Ward identity from the generator of translations $L_{-1}$ imposes dependence on differences $z_{i}-z_{j}$. Supertranslation operators $G^{\pm,H}_{-\frac{1}{2}}$ further impose dependence on superdifferences
\begin{eqnarray}\label{zij}
 \mathsf{Z}_{ij}&=&z_{i}-z_{j}+\frac{1}{8}\left(\theta^{-}_{i}\theta_{j}^{+}+\theta_{i}^{+}\theta_{j}^{-}\right)+\theta^{H}_{i}\theta^{H}_{j},\\
\label{thetaij}
 \theta_{ij}^{\pm, H}&=&\theta_{i}^{\pm,H}-\theta_{j}^{\pm, H}.
\end{eqnarray}

Writing down other Ward identities explicitly and transforming them to new coordinates can become tedious for high values of $n$. Fortunately, this turns out to be unnecessary by using the product ansatz
\begin{equation}\label{2pf}
\begin{array}{c}
\langle \Phi_{h_{1}, t_{1}, q_{1}}\left(\mathsf{Z}_{1}\right)...\Phi_{h_{n}, t_{n}, q_{n}}\left(\mathsf{Z}_{n}\right)\rangle=
\Upsilon_{h_{1},...,h_{n}}\left(\mathsf{Z}_{12},...,\mathsf{Z}_{n-1,n}\right)\Omega_{t_{1},q_{1};...;t_{n}q_{n}}\left(\mathsf{Z}_{1},...\mathsf{Z}_{n}\right).
\end{array}
\end{equation}
While $\Upsilon_{h_{1},...,h_{n}}$ is a function of (\ref{zij}) and satisfies the action of the conformal group (\ref{L-1})-(\ref{L1}), $\Omega_{t_{1},q_{1};...;t_{n}q_{n}}$ depends on both (\ref{zij}) and (\ref{thetaij}). For correlation functions of $n$ fields, the parameterization of $\frac{n\left(n-1\right)}{2}\times 4$ superdifferences is chosen such that $i<j$. Using (\ref{L1}), the solution to $\Upsilon_{h_{1},...,h_{n}}$ can be found to be
\begin{eqnarray}\label{sigma}
\Upsilon_{h_{1},...,h_{n}}\left(\mathsf{Z}_{12},...,\mathsf{Z}_{n-1,n}\right)&=&\left\{ \begin{array}{ll}
n=2: &  \mathsf{Z}_{12}^{-2h_{1}}\delta_{h_{1},h_{2}} \\
n>2: & F\left(x_{1}, ...,x_{n-3} \right)\displaystyle \prod_{i<j}\mathsf{Z}_{ij}^{-\Delta_{ij}}\\
\end{array} \right.
\\
  \displaystyle \sum_{i,j;i<j}\Delta_{i,j}&=&\sum_{i}h_{i}, \;\;\; \sum_{j;i<j}\Delta_{ij}=2h_{i}.\nonumber
\end{eqnarray} 
This solution is the same as in N=0,1,2 theories, up to a redefinition of $\mathsf{Z}_{ij}$
\begin{center}
\begin{tabular}{lr}
$\mathsf{Z}_{ij}=z_{i}-z_{j}$\;\;\; &$ \textnormal{for N=0}$,  \\
$\mathsf{Z}_{ij}=z_{i}-z_{j}-\theta_{i}\theta_{j}$\;\;\; &$ \textnormal{for N=1}$,  \\
$\mathsf{Z}_{ij}=z_{i}-z_{j}-\theta_{i}^{+}\theta_{j}^{-}-\theta_{i}^{-}\theta_{j}^{+}$\;\;\; &  $\textnormal{for N=2}$.  \\
\end{tabular}
\end{center}
The function $\Upsilon_{h_{1},...,h_{n}}\left(\mathsf{Z}_{12},...,\mathsf{Z}_{n-1,n}\right)$ is expected to remain valid in the ansatz (\ref{2pf}) for N=4 super algebra generated by OSp(2$\vert$4) for supertranslationally invariant superdifferences  $\mathsf{Z}_{ij}$. The function $F(x_{1}, ...,x_{n-3} )$ depends on $n-3$ independent cross-ratios. Additional constraints on the function $F$ can be obtained from singular vectors in Verma modules. In case of logarithmic dependence \cite{Gurarie}\cite{Flohr:2001zs} the representations in the correlation function are indecomposable with respect to $L_{0}$. One can compute correlation functions of indecomposable representations from (\ref{sigma}) using the ``derivation trick"  \cite{Khorrami:1998kw}\cite{RahimiTabar:1996ub}.

In N=1 theory, the function $\Omega$ is given by $1$ for the two-point function and by increasingly longer expressions with growing $n$ \cite{Fuchs:1986ew}. 
\newpage
In the case of the N=2 $n$-point function, $\Omega$ is given by the simple, closed form \cite{Kiritsis:1987np}
\begin{equation*}
 \Omega=\exp \left( \displaystyle \sum_{i<j}^{n} A_{ij} \frac{\theta_{ij}^{+}\theta_{ij}^{-}}{\mathsf{Z}_{ij}} \right)\delta_{\sum_{i=1}^{n}q_{i},0}
\end{equation*}
\begin{equation*}
 \displaystyle \sum_{j=1, i\neq j}^{n}A_{ij}=-q_{i}, \;\;\; A_{ij}=-A_{ji},
\end{equation*}
where $q_{i}$ is the eigenvalue to the U(1)-isospin operator.

In N=3 theory, the form of $\Omega$ depends on the value of the isospin. In the following, we will derive  $\Omega_{t_{1},q_{1};...;t_{n}q_{n}}$.

The Ward identities arising from the action of $T^{\pm,H}_{0}$ on $\Omega_{t_{1},q_{1};...;t_{n}q_{n}}$ can be rewritten using classical operators $t^{\pm,H}_{0}$ (\ref{cl}). The $n$-point function, up to normalization, is an eigenstate of su(2) operators $t_{0}^{\pm H(1...n)}$
\begin{equation}\label{tomega}
\begin{array}{rcl}

t_{0}^{H\left(1...n\right)}\Omega_{q_{1},...,q_{n}}&=&\left( \displaystyle \sum_{i<j}\left(\theta_{ij}^{+}\partial_{\theta_{ij}^{+}}-\theta_{ij}^{-}\partial_{\theta^{-}_{ij}}\right)\right)\Omega_{q_{1},...,q_{n}}=\displaystyle \left( \sum_{i=1}^{n}q_{i}\right) \Omega_{t_{1},q_{1};...;t_{n},q_{n}}\\

t_{0}^{\pm\left(1...n\right)}\Omega_{q_{1},...,q_{n}}&=&\left(\displaystyle \sum_{i<j}\left(\pm \frac{1}{2}\theta_{ij}^{\pm}\partial_{\theta_{ij}^{H}}\mp 4 \theta_{ij}^{H}\partial_{\theta^{\mp}_{ij}}\right)\right)\Omega_{q_{1},...,q_{n}}\\
 &=& \Omega_{t_{1},q_{1}\pm 1;...;t_{n},q_{n}}+...+\Omega_{t_{1},q_{1};...;t_{n},q_{n}\pm 1}.
\end{array}
\end{equation}

Considering the special case of the two-point function, we see that the possible superspace components of the two-point function are just (\ref{su2basis}) with an implied position index on all Grassmann variables. It will be seen that $L_{1}$ does not allow terms with an uneven number of Grassmann variables. We immediately conclude
\[
\Omega_{0,0;0,0}=1.
\]

For $q_{1}=t_{1}$, $q_{2}=t_{2}$, $t_{1}+t_{2}\neq 0$, (\ref{tomega}) returns $t_{1}+t_{2}=1$. There are two solutions,  $t_{1}=t_{2}=\frac{1}{2}$ and $t_{1}=1$, $t_{2}=0$. The latter ansatz does not satisfy the remaining Ward identities. We are left with the result that the $t=1$ superfunction representation of OSp(2$\vert$3) is just $\Omega_{\frac{1}{2},\frac{1}{2};\frac{1}{2},\frac{1}{2}}$. All representations with $t\neq 0$, $t \neq \frac{1}{2}$ decouple from the theory since they do not correlate with any other field, including themselves. This result is consistent with \cite{SchwimmerSeiberg}, where states $h=0$, $q=0$ and $h=\frac{1}{4}$, $q=\frac{1}{2}$ were identified in the $c=\frac{3}{2}$ theory.

Since there are only $t=0$ and $t=\frac{1}{2}$ primary fields in the N=3 theory, we use the short-hand notation
\[
\Omega_{0,0;0,0}\equiv\Omega_{00}, \;\;\;\;\;\; \Omega_{\frac{1}{2},\frac{1}{2};\frac{1}{2},\frac{1}{2}}\equiv \Omega_{++}, \;\;\;\;\;\; \Omega_{\frac{1}{2},-\frac{1}{2};\frac{1}{2},-\frac{1}{2}}\equiv \Omega_{--}.
\]
\[
\Omega_{\frac{1}{2},\frac{1}{2};\frac{1}{2},-\frac{1}{2}}\equiv \Omega_{+-} \;\;\;\;\;\; \Omega_{\frac{1}{2},-\frac{1}{2};\frac{1}{2},\frac{1}{2}}\equiv \Omega_{-+}.
\]

To obtain further constraints, one needs to use the remaining five Ward identities. The action of $L_{0}$ on the two-point function is
\begin{equation}
  \mathcal{L}_{0}^{(12)}F_{2}=\left(h_{1}+h_{2}+\mathsf{Z}_{12}\partial_{\mathsf{Z}_{12}}+\\ 
\frac{1}{2}\left(\theta_{12}^{+}\partial_{\theta_{12}^{+}}+\theta_{12}^{-}\partial_{\theta_{12}^{-}}+\theta_{12}^{H}\partial_{\theta_{12}^{H}}\right)\right)F_{2}=0.
\end{equation}
Besides imposing similar conformal conditions as in the N=0 theory, it scales all terms containing nilpotent variables to the right dimension. Note that $\Omega$ is dimensionless. 

The map from superspace points $\theta_{i}$, $z_{i}$ to superdifferences $\theta_{ij}^{\pm, H}$, $\mathsf{Z}_{ij}$ is a projection. Because of that, we have to introduce new combinations of superspace coordinates $\xi_{ij}^{\pm, H}$, $W_{ij}$ (``supersums'') to express down Ward identities for $L_{1}$ and $G_{\frac{1}{2}}^{\pm, H}$,
\begin{eqnarray*}
 \xi_{ij}^{\pm, H}&=&\theta_{i}^{\pm,H}+\theta_{j}^{\pm, H},\\
 \mathsf{W}_{ij}&=&z_{i}+z_{j}+\frac{1}{8}\left(\theta^{-}_{i}\theta_{j}^{+}+\theta_{i}^{+}\theta_{j}^{-}\right)+\theta^{H}_{i}\theta^{H}_{j}.
\end{eqnarray*}
The remaining four Ward identities are the most difficult to compute and understand intuitively. They arise from generators of special and special superconformal transformations which mix bosonic and fermionic coordinates in a non-trivial way. The most computationally difficult terms are  of the form  $\mathsf{Z}\partial_{\theta}$. The desired Ward identities are
\small \begin{equation}\label{L1WI}
\begin{array}{rcl}
\left(\mathcal{L}_{1}^{\left(1\right)}+\mathcal{L}_{1}^{\left(2\right)}\right)F_{2}&=& 
\Big( h_{1}\left(\mathsf{W}_{12}+\mathsf{Z}_{12}-\frac{1}{8}\left(\theta_{12}^{-}\xi_{12}^{+}+\theta_{12}^{+}\xi_{12}^{-}\right)-\theta_{12}^{H}\xi_{12}^{H}\right)+h_{2}\left(\mathsf{W}_{12}-\mathsf{Z}_{12}\right) \\
& & {}+ 
\frac{1}{16}\left(\xi_{12}^{+}+\theta_{12}^{+}\right)\left(\xi_{12}^{H}+\theta_{12}^{H}\right)J_{1}^{-}
+\frac{1}{16}\left(\xi_{12}^{+}-\theta_{12}^{+}\right)\left(\xi_{12}^{H}-\theta_{12}^{H}\right)J_{2}^{-} \\ 
& & {}-
\frac{1}{16}\left(\xi_{12}^{-}+\theta_{12}^{-}\right)\left(\xi_{12}^{H}+\theta_{12}^{H}\right)J_{1}^{+}
-\frac{1}{16}\left(\xi_{12}^{-}-\theta_{12}^{-}\right)\left(\xi_{12}^{H}-\theta_{12}^{H}\right)J_{2}^{+} \\ 
& & {}+ 
\frac{1}{32}\left(\xi_{12}^{+}+\theta_{12}^{+}\right)\left(\xi_{12}^{-}+\theta_{12}^{-}\right)J_{1}^{H}
+\frac{1}{32}\left(\xi_{12}^{+}-\theta_{12}^{+}\right)\left(\xi_{12}^{-}-\theta_{12}^{-}\right)J_{2}^{H} \\ 
& & {}+ 
\left(\mathsf{W}_{12}\mathsf{Z}_{12}-\frac{1}{16}\mathsf{Z}_{12}\left(\theta_{12}^{-}\xi_{12}^{+}+\theta_{12}^{+}\xi_{12}^{-}\right)-\frac{1}{2}\mathsf{Z}_{12}\theta_{12}^{H}\xi_{12}^{H}\right)\partial_{\mathsf{Z}_{12}} \\ 
& & {}+ 
\frac{1}{2}\big(\mathsf{Z}_{12}\xi_{12}^{+}+\mathsf{W}_{12}\theta_{12}^{+}+\frac{1}{16}\theta_{12}^{+}\xi_{12}^{+}\left(\xi_{12}^{-}+\theta_{12}^{-}\right)
\\ & & {}-\frac{1}{2}\theta_{12}^{H}\xi_{12}^{H}\left(\xi_{12}^{+}+\theta_{12}^{+}\right)\big)\partial_{\theta_{12}^{+}}+ 

\frac{1}{2}\big(\mathsf{Z}_{12}\xi_{12}^{-}+\mathsf{W}_{12}\theta_{12}^{-}

\\& &{}+\frac{1}{16}\theta_{12}^{-}\xi_{12}^{-}\left(\xi_{12}^{+}+\theta_{12}^{+}\right)
-\frac{1}{2}\theta_{12}^{H}\xi_{12}^{H}\left(\xi_{12}^{-}+\theta_{12}^{-}\right)\big)\partial_{\theta_{12}^{-}} \\ 
& & {}+ 
\frac{1}{2}\left(\mathsf{Z}_{12}\xi_{12}^{H}+\mathsf{W}_{12}\theta_{12}^{H}-\frac{1}{16}\left(\theta_{12}^{-}\xi_{12}^{+}+\theta_{12}^{+}\xi_{12}^{-}\right)\left(\xi_{12}^{H}+\theta_{12}^{H}\right)\right)\partial_{\theta_{12}^{H}}\Big)F_{2}  =  0,
\end{array}
\end{equation}
\begin{equation}\label{GH12WI}
  \begin{array}{rcl}
\left(\mathcal{G}^{H\left(1\right)}_{\frac{1}{2}}+\mathcal{G}^{H\left(2\right)}_{\frac{1}{2}}\right)F_{2}&=&\Big( -4\left(\left(h_{1}+h_{2}\right)\xi_{12}^{H}+\left(h_{1}-h_{2}\right)\theta_{12}^{H}\right) \\
& & {}-
\frac{1}{2}\left(\xi_{12}^{+}+\theta_{12}^{+}\right)J_{1}^{-}-\frac{1}{2}\left(\xi_{12}^{+}-\theta_{12}^{+}\right)J_{2}^{-}
\\
& & {}+\frac{1}{2}\left(\xi_{12}^{-}+\theta_{12}^{-}\right)J_{1}^{+}+\frac{1}{2}\left(\xi_{12}^{-}-\theta_{12}^{-}\right)J_{2}^{+} \\
& & {}-
4 \mathsf{Z}_{12}\xi_{12}^{H}\partial_{\mathsf{Z}_{12}}-
2\left(\xi_{12}^{H}\theta_{12}^{-}+\theta_{12}^{H}\xi_{12}^{-}\right)\partial_{\theta_{12}^{-}}-2\left(\xi_{12}^{H}\theta_{12}^{+}+\theta_{12}^{H}\xi_{12}^{+}\right)\partial_{\theta_{12}^{+}} \\ 
& & {}-
4\left(\mathsf{Z}_{12}-\frac{1}{16}\left(\theta_{12}^{-}\xi_{12}^{+}+\theta_{12}^{+}\xi_{12}^{-}\right)-\frac{1}{2}\theta_{12}^{H}\xi_{12}^{H}\right)\partial_{\theta_{12}^{H}}\Big)F_{2} = 0,
 \end{array}
\end{equation}
\begin{equation}\label{GPM12WI}
 \begin{array}{rcl}
\left(\mathcal{G}^{\pm \left(1\right)}_{\frac{1}{2}}+\mathcal{G}^{\pm \left(2\right)}_{\frac{1}{2}}\right)F_{2}&=&\Big(\pm h_{1}\left(\xi_{12}^{\pm}+\theta_{12}^{\pm}\right)\pm h_{2}\left(\xi_{12}^{\pm}-\theta_{12}^{\pm}\right) \\
& & {}+
\left(\xi_{12}^{H}+\theta_{12}^{H}\right)J_{1}^{\pm}+\left(\xi_{12}^{H}-\theta_{12}^{H}\right)J_{2}^{\pm} \\
& & {}+
\frac{1}{2}\left(\xi_{12}^{\pm}+\theta_{12}^{\pm}\right)J_{1}^{H}+\frac{1}{2}\left(\xi_{12}^{\pm}-\theta_{12}^{\pm}\right)J_{2}^{H}\pm\xi_{12}^{\pm}\mathsf{Z}_{12}\partial_{\mathsf{Z}_{12}} \\ 
& & {}+
\left(\pm8\mathsf{Z}_{12}\mp\theta_{12}^{\mp}\xi_{12}^{\pm}-4\theta_{12}^{H}\xi_{12}^{H}\right)\partial_{\theta_{12}^{\mp}}
\\
&&{}\pm\frac{1}{2}\left(\theta_{12}^{\pm}\xi_{12}^{H}+\xi_{12}^{\pm}\theta_{12}^{H}\right)\partial_{\theta_{12}^{H}}\Big)F_{2} = 0.
 \end{array}
\end{equation}
\normalsize
Since $G^{\pm,H}_{\frac{1}{2}}$ is generated by $L_{1}$ and $G^{\pm,H}_{-\frac{1}{2}}$, we do not have to account for their action if the remaining Ward identities are satisfied.\footnote[1]
{We should note that there also exists a quadratic Casimir operator $T_{0}^{C}$ of the theory containing double derivatives with respect to nilpotent variables. This operator turns out to be of no particular use for determining correlation functions. We show how it works for correlation functions of primary fields for the simple case $T^{\pm}_{0}\vert h, \pm q \rangle =0$ (equivalently, $q=\pm t$). We compute

\[
T_{0}^{C}=T_{0}^{\mp}T_{0}^{\pm}+\left(T_{0}^{H}\right)^{2}\pm T_{0}^{H}
\]
through the two-point function. Since the  first term disappears and the second one returns an already known identity, we obtain the superdifferential equation
\begin{equation*} \label{T2WI}
\begin{array}{c}
 \left(\left(\mathcal{T}_{0}^{H(1)}\right)^{2}+2\mathcal{T}_{0}^{H(1)}\mathcal{T}_{0}^{H(2)}+\left(\mathcal{T}_{0}^{H(2)}\right)^{2}\right)F_{2}=\\
 \left(-2\theta^{+}_{12}\theta_{12}^{-}\partial_{\theta_{12}^{-}}\partial_{\theta_{12}^{+}} 
+\left(1-2\left(q_{1}+q_{2}\right)\right)\theta_{12}^{+}\partial_{\theta_{12}^{+}}+\left(1+2\left(q_{1}+q_{2}\right)\right)\theta_{12}^{-}\partial_{\theta_{12}^{-}}+\left(q_{1}+q_{2}\right)^{2}\right)F_{2}=0.
\end{array}
\end{equation*}
}

It is now easy to determine all two-point functions. Using (\ref{L1WI}) we find that the terms with an uneven number of Grassmann variables are not contained in the two-point function. We are left with
\begin{equation*}
\Omega_{++}  \sim  \displaystyle \frac{\theta_{12}^{+}\theta^{H}_{12}}{\mathsf{Z}_{12}},\ \ \ \ \
\Omega_{+-}+\Omega_{-+}  \sim  \displaystyle \frac{\theta_{12}^{+}\theta^{-}_{12}}{\mathsf{Z}_{12}}\ \ \ \ \ \mathrm{and}\ \ \ \ \
\Omega_{--}  \sim  \displaystyle \frac{\theta_{12}^{-}\theta^{H}_{12}}{\mathsf{Z}_{12}}.
\end{equation*}
Using equations (\ref{G12H}), (\ref{G12PM}) we obtain conditions for mixed correlation functions
\begin{equation*}
\Omega_{-+} +\Omega_{+-} = \displaystyle \frac{\theta_{12}^{-}\theta_{12}^{+}}{2\mathsf{Z}_{12}}\ \ \ \ \ \mathrm{and}\ \ \ \ \
\Omega_{-+} - \Omega_{+-} = 8.
\end{equation*}
Then the complete set of two-point correlation functions, up to a constant, is given by
\begin{equation}\label{omega}
\Omega_{00} = 1,\ \ \ \ \ 
\begin{array}{rclcrcl}
\displaystyle \Omega_{++} &=& \displaystyle  \frac{\theta^{+}_{ij}\theta^{H}_{12}}{\mathsf{Z}_{12}}, & &
\displaystyle   \Omega_{-+} &=& \displaystyle  \frac{\theta_{12}^{-}\theta_{12}^{+}}{4\mathsf{Z}_{12}} +4, \\
\displaystyle \Omega_{--} &=& \displaystyle  \frac{\theta^{-}_{12}\theta^{H}_{12}}{\mathsf{Z}_{12}}, & &
\displaystyle  \Omega_{+-} &=& \displaystyle  \frac{\theta_{12}^{-}\theta_{12}^{+}}{4\mathsf{Z}_{12}}-4. \\
\end{array}
\end{equation}
The solutions $\Omega_{00}$ and $\Omega_{++}$ have been found in \cite{Nagi:2003pb}. Determining the normalization, we find that the $j=1$ superfunction eigenstates
\begin{equation}\label{fu}
\begin{array}{rcl}
t^{H (12)} \Omega_{1,q} & = & q \Omega_{1,q}, \\
t^{\pm (12)}\Omega_{1,q} & = & \Omega_{1,q \pm 1}
\end{array}
\end{equation}
are given by
\begin{equation}\label{1rep}
\begin{array}{rcl}
\Omega_{t=1,q=1}  &=& \Omega_{++},\\
\Omega_{t=1,q=0}  &=& \frac{1}{2}\left(\Omega_{+-}+\Omega_{-+}\right),\\ 
\Omega_{t=1,q=-1} &=& \Omega_{--}.\\
\end{array}
\end{equation}
This can be generalized for any $t \in \mathbb{N}$, which will be shown in the next section.

\section{The $n$-Point Functions}
\noindent Solving the $n$-point function amounts to finding all $\Omega_{q_{1},...,q_{n}}$. We will assume $q_{1}, q_{2} \in \lbrace \pm \frac{1}{2} \rbrace $, since $q=0$ fields have no effect on $\Omega$, thus acting as identity primiary fields,
\[
\Omega_{q_{1},  ...q_{n}}=\Omega_{0, q_{1},..., q_{n}}=\Omega_{q_{i}, 0,...,  q_{n}}=\ldots .
\]

Under this condition, there are no solutions for odd $n$, since there are only eigenfunctions corresponding to integer eigenvalues to the differential operator (\ref{tomega}). The only non-trivial three-point functions are two-point functions with an inserted isospin singlet.
Representations transforming under operators (\ref{tomega}) are constructed from products of nilpotent variables. For a given product $ \theta_{i_{1}j_{1}}^{\pm,H}...\theta_{i_{\frac{n}{2}}j_{\frac{n}{2}}}^{\pm,H}$, $n$ indices have to be distributed without appearing twice. By assigning index positions, we choose a particular pairwise contraction of fields in the correlation function, all of which lead to equivalent solutions. Since the Ward identities in $\Omega$ are linear, in contrast to the complete correlation functions (\ref{2pf}), they are just sums of conditions for the $\frac{n}{2}$ two-point functions. There are no additional solutions, as can be seen by explicit calculation. Thus, every contraction returns one of the terms in (\ref{omega}) and any solution to an $n$-point function can be expressed as a linear combination of all possible contractions,
\[
\Omega_{q_{1},...,q_{n}}=\displaystyle \sum_{\textnormal{Contractions}}c_{k}\prod_{i_{k}<j_{k}}\Omega^{i_{k}j_{k}}_{q_{i_{k}}q_{j_{k}}},
\]
where
$i_{1}\neq ...\neq i_{\frac{n}{2}}\neq j_{1} \neq...\neq j_{\frac{n}{2}}$ are
all mutually distinct.

This result is just an explicit realization of the crossing symmetry. We will restrict our considerations to one particular contraction, for example $(1,2)(3,4)...(n-1,n)$ and dismiss indices on $i$, $j$ for the sake of keeping a cleaner notation
\[
\Omega_{q_{1},  ...q_{n}}=\prod_{ i<j}\Omega^{ij}_{q_{i},q_{j}}=\Omega^{12}_{q_{1},q_{1}}\Omega^{34}_{q_{3},q_{4}}...\Omega^{n-1,n}_{q_{n},q_{n-1}}.
\]
The two upper indices on $\Omega_{q_{1}, q_{2}}$ in (\ref{omega}) correspond to the ``position" in superspace and replace the double index ``$12$'' in (\ref{omega}).

We show how this works for $n=4$. Consider $\Omega_{++++}$ and choose the particular contraction $(1,2)(3,4)$. This is the $t=2,q=2$ representation, from which we can determine representations with lowered $q$. At the same time,  (\ref{1rep}) can be used to reexpress $\Omega_{2,q}$ in terms of $\Omega_{1,q}$, 
\begin{equation*}
\begin{array}{rcl}
\Omega_{2,2}&=&\Omega_{++}^{12}\Omega_{++}^{34}\\
		&=&\Omega_{1,1}^{12}\Omega_{1,1}^{34}\\	
\Omega_{2,1}&=&\displaystyle \frac{1}{2}
(\Omega_{+-}^{12}\Omega_{++}^{34}+\Omega_{-+}^{12}\Omega_{++}^{34}+\Omega_{++}^{12}\Omega_{-+}^{34}+\Omega_{++}^{12}\Omega_{+-}^{34})\\
		&=&\displaystyle \frac{1}{2}(\Omega_{1,1}^{12}\Omega_{1,0}^{34}+\Omega_{1,0}^{12}\Omega_{1,1}^{34}) \\
\Omega_{2,0}&=&\displaystyle \frac{1}{4}\left(\Omega_{--}^{12}\Omega_{++}^{34}+\Omega_{+-}^{12}\Omega_{-+}^{34}+ \Omega_{+-}^{12}\Omega_{+-}^{34}+  \Omega_{-+}^{12}\Omega_{-+}^{34}+\Omega_{-+}^{12}\Omega_{+-}^{34}+ \Omega_{++}^{12}\Omega_{--}^{34} \right) \\
	&=&\displaystyle  \frac{1}{4}(\Omega_{1,1}^{12}\Omega_{1,-1}^{34}+2\Omega_{1,0}^{12}\Omega_{1,0}^{34}+\Omega_{1,-1}^{12}\Omega_{1,1}^{34}) \\
\Omega_{2,-1}&=&\displaystyle \frac{1}{2}(\Omega_{-+}^{12}\Omega_{--}^{34}+\Omega_{+-}^{12}\Omega_{--}^{34}+\Omega_{--}^{12}\Omega_{+-}^{34}+\Omega_{--}^{12}\Omega_{-+}^{34})\\
&=&\displaystyle  \frac{1}{2}(\Omega_{1,-1}^{12}\Omega_{1,0}^{34}+\Omega_{1,0}^{12}\Omega_{1,-1}^{34})\\
\Omega_{2,-2} &=&\Omega_{--}^{12}\Omega_{--}^{34}\\
&=&\Omega_{1,-1}^{12}\Omega_{1,-1}^{34}.
\end{array}
\end{equation*}

All there is left to do is generalizing (\ref{fu}) and (\ref{1rep}). Performing the same calculation that led to (\ref{su2basis}) with 3$\times (n-1)$ independent variables, we find that there is a basis  $\Omega_{t,q}$ of a superfunction so that it is a representation  $t=\frac{n}{2}$,
\begin{equation}\label{to}
t^{H\left(1...n\right)}\Omega_{t,q}=q \Omega_{t,q}\ \ \ \ \ \mathrm{and}\ \ \ \ \
t^{\pm\left(1...n\right)}\Omega_{t,q}=\Omega_{t,q\pm1}.
\end{equation}
The $q=t$ representation is given by the correlation function of $n=2t$ isospin $\frac{1}{2}$ fields. For an arbitrary $q$, there are $\binom{2t}{ t-\vert q \vert }$ solutions to the equation
\begin{equation}\label{sumq}
\sum_i^{n} q_{i}=q, \;\;\;\; q_{i} \in \lbrace \pm \frac{1}{2} \rbrace , \;\;\;\; q\in \mathbb{N}_{0}.
\end{equation}
This is the number of ways to distribute $ t-\vert q \vert $ elements (fields) on a set of $n=2t$ elements ($n$-point correlation function). Thanks to our short-hand notation, the general solution to (\ref{to}) can be written as
\begin{equation}
\Omega_{t,q}= \left( \frac{1}{2} \right)^{t-\vert q \vert } \displaystyle \sum_{\sum q_{i}=q  }\prod_{i<j }\Omega^{ij}_{q_{i},q_{j}}. 
\end{equation}
The sum is understood to be taken over solutions to  (\ref{sumq}). The factors $\frac{1}{2}$ are a consequence of the normalization (\ref{norm}).

\section{Conclusion}
\noindent In the presented paper we have shown how the super Ward identities can be solved in the Neveu-Schwarz sector of the N=3 superconformal field theory. This is done by using conditions arising from the action of the supercurrent on correlation functions to explicitly construct su(2) representations in superspace.  

An interesting consequence is that the isospin representation content of the field theory is quite small, consisting only of  $t=0$ and $t=\frac{1}{2}$ representations. This should make computation of various quantities of the theory like the Kac determinant easier than previously assumed. We are confident that a similar result can be obtained for the Ramond algebra of global transformations. This is supported by results in \cite{SchwimmerSeiberg}, where a $h=\frac{1}{16}$, $q=0$ and a  $h=\frac{5}{16}$, $q=\frac{1}{2}$ representation were found in the Ramond sector.  

In principle, similar techniques can be applied for supersymmetric extensions with N$>$3. The case N=4, where the so(4) superconformal current can be mapped to su(2)$\times$su(2) seems to be particularly feasible. It seems likely that representation spaces of higher-dimensional extensions can be found by generalizing our method. Recent  results on orthosymplectic groups and their representations \cite{Fre:2009ki}\cite{Luo:2010si} might be elucidating in the process of gaining a deeper understanding of superconformal theories.

\section*{Acknowledgements}
\noindent The authors are grateful to Jasbir Nagi and Elias Kiritsis for comments and helpful discussions over e-mail.

\bibliographystyle{elsarticle-num}

\end{document}